\documentclass[%
preprint,
superscriptaddress,
amsmath,amssymb,
]{revtex4-2}

\usepackage{graphicx}
\usepackage{dcolumn}
\usepackage{bm}
\usepackage{color}
\usepackage{xfrac}
\usepackage[normalem]{ulem}
\usepackage{cancel}
\usepackage{url}
\usepackage{mathtools}
\usepackage{hyperref}
\usepackage{caption}
\usepackage{physics}
\usepackage{xcolor}
\usepackage{comment}
\usepackage{ulem}
\usepackage{xspace}
\usepackage[version=3]{mhchem} 

\newcommand{\ai}{\textit{ab initio}\xspace}
\newcommand{\lumen}{\textit{Lumen}\xspace}
\newcommand{\sn}{\ce{Si3N4}\xspace} 
\newcommand{\mos}{\ce{MoS2}\xspace} 
\newcommand{\ima}{$\operatorname{Im}\left\{\tilde{\alpha}\right\}$\xspace} 
\newcommand{\ime}{$\operatorname{Im}\left\{\varepsilon_r\right\}$\xspace} 

\begin{document}
\title{Extreme-ultraviolet optical response of atomically-thin molybdenum disulfide}

\author{G. Fiorentini}
\thanks{These authors contributed equally to this work}
\author{N. Di Palo}
\thanks{These authors contributed equally to this work}
\author{G. Inzani}
\author{G. L. Dolso}
\author{S. Bonetti}
\affiliation{Department of Physics, Politecnico di Milano, Milano, 20133, Italy}

\author{Q. Li}
\affiliation{Department of Chemistry, Columbia University, New York, NY 10027, USA}
\author{F. Liu}
\affiliation{Department of Chemistry, Columbia University, New York, NY 10027, USA}
\affiliation{Department of Chemistry, Stanford University, Stanford, CA 94305, USA}
\author{X. Zhu}
\affiliation{Department of Chemistry, Columbia University, New York, NY 10027, USA}

\author{A. Giglia}
\author{N. Mahne}
\affiliation{CNR-IOM – Istituto Officina dei Materiali, National Research Council of Italy, Trieste, 34149, Italy}
\affiliation{INAF - Osservatorio Astrofisico di Torino, Torino, Italy}
\author{L. Pasquali}
\affiliation{CNR-IOM – Istituto Officina dei Materiali, National Research Council of Italy, Trieste, 34149, Italy}
\affiliation{Dipartimento di Ingegneria E.Ferrari, Università di Modena e Reggio Emilia, Modena, 41125, Italy}
\affiliation{Department of Physics, University of Johannesburg, Auckland Park 2006, South Africa}

\author{M. D’Alessandro}
\affiliation{Istituto di Struttura della Materia-CNR (ISM-CNR), Roma, 00133, Italy}
\author{M. Malakhov}
\affiliation{Institute of Metal Physics of the Ural Branch of the Russian Academy of Sciences, Yekaterinburg, 620108, Russia}
\author{M. Camarasa-G\'omez}
\affiliation{Centro de Física de Materiales (CFM-MPC), CSIC-UPV/EHU, Donostia-San Sebastián, 20018, Spain}
\author{J. J. Esteve-Paredes}
\affiliation{Departamento de Física de la Materia Condensada, Universidad Autónoma de Madrid, Madrid, 28049, Spain}
\author{J. J. Palacios}
\affiliation{Departamento de Física de la Materia Condensada, Universidad Autónoma de Madrid, Madrid, 28049, Spain}
\author{R. Borrego-Varillas}
\affiliation{Institute for Photonics and Nanotechnologies, IFN-CNR, Milano, 20133, Italy}
\author{M. Nisoli}
\affiliation{Department of Physics, Politecnico di Milano, Milano, 20133, Italy}
\affiliation{Institute for Photonics and Nanotechnologies, IFN-CNR, Milano, 20133, Italy}
\author{A. Pic\'on}
\affiliation{Instituto de Ciencia de Materiales de Madrid (ICMM-CSIC), Madrid, 28049, Spain}
\author{D. Sangalli}
\affiliation{Istituto di Struttura della Materia-CNR (ISM-CNR), Milano, 20133, Italy}
\author{M. Lucchini}
\affiliation{Department of Physics, Politecnico di Milano, Milano, 20133, Italy}
\affiliation{Institute for Photonics and Nanotechnologies, IFN-CNR, Milano, 20133, Italy}
\email{matteo.lucchini@polimi.it}

\begin{abstract}
    We report multi-angle reflectivity measurements in the extreme-ultraviolet (XUV) range for mono- and bilayer \mos on a \sn substrate. Using a single-sheet 2D conductivity model, we extract the complex optical response of the \mos bilayer between 25 and 90\,eV and derive an effective refractive index by introducing a thickness equal to the interlayer spacing. The \mos monolayer response is consistently reproduced either by halving the 2D conductivity or the effective thickness, indicating a robust scaling with layer number. The resulting optical constants display a broad resonance at the Mo N$_{2,3}$ edge with no signatures of sharp core-exciton features despite the reduced dimensionality. First-principles calculations reproduce the experimental results and show that local-field (Hartree) effects dominate the XUV response, while screened-exchange (SEX) contributions remain weak and mainly induce spectral shifts. Our analysis demonstrates that excitonic effects play a minor role in the XUV optical response of atomically thin \mos, highlighting key differences with respect to the visible and infrared regimes, and calling for a reassessment of the use of Mo-based transition metal dichalcogenides in attosecond spectroscopy and XUV excitonics.
\end{abstract}

\maketitle

\section{Introduction}

\begin{figure}[htbp]
	\begin{center}
		\begin{tabular}{c}
			\includegraphics[width=0.47\textwidth]{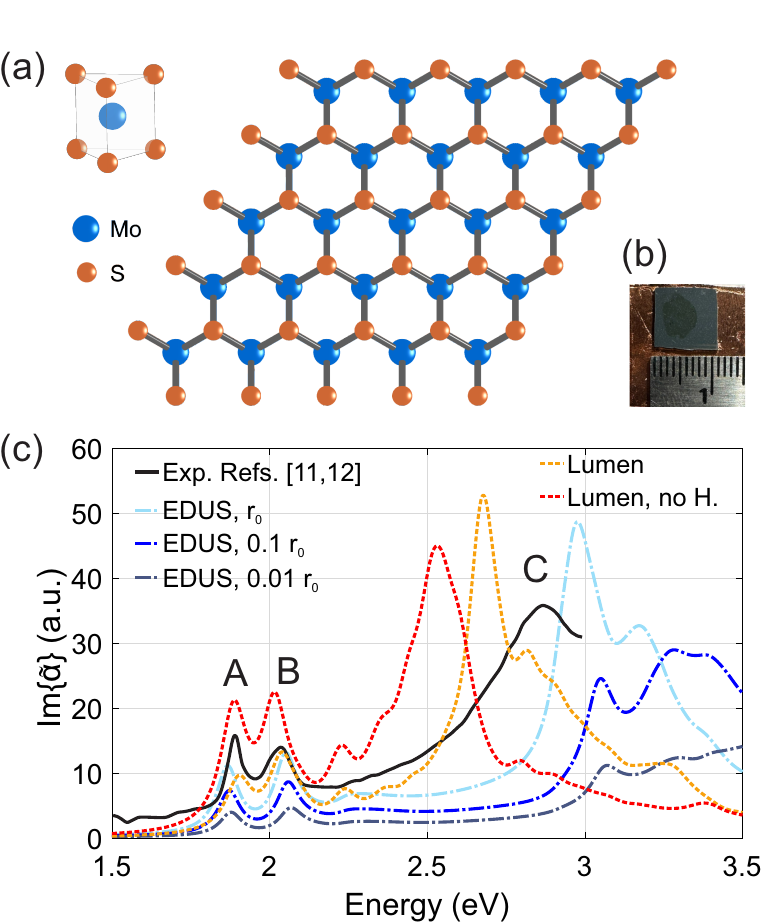}
		\end{tabular}
	\end{center}
	\caption[example]{\label{Fig:1}
		\textbf{(a)} \mos unit cell (left) and top view of the crystal structure (right). \textbf{(b)} Optical image of the 2L \mos sample deposited on a \sn substrate used in the experiment. \textbf{(c)} Imaginary part of the complex polarizability, \ima, as a function of energy. The black solid curve shows the experimental values from Refs.~\cite{Mak2010,Qiu2013}. The dot-dashed blue curves are calculated with EDUS while varying the screening of the Hartree term (see Sec.~\ref{Sec:EDUS}). The dotted curves are calculated with \lumen (see Sec.~\ref{Sec:Lumen}) with (orange) and without (red) inclusion of the Hartree term. To enable a direct comparison of the excitonic structures (labeled A, B, and C), the theoretical spectra have been rigidly shifted in energy to match the experimental values below 2.2\,eV. Since the calculations were performed at 0~K, this procedure partially accounts for the expected temperature-dependent shift of the absorption spectrum, although it cannot reproduce the variation in relative spacing between the A/B and C peaks \cite{Molina-Snchez2016}.
	}
\end{figure}

The optical response of atomically-thin transition metal dichalcogenides (TMDs) is dominated by excitonic effects arising from reduced dielectric screening and enhanced Coulomb interactions in two dimensions \cite{Moody2016,Wang2018}. As a consequence, monolayer TMDs exhibit tightly bound excitons with large binding energies and pronounced spectral signatures that govern their optical properties in the visible and near-infrared (IR) spectral ranges~\cite{Li2014} (see Fig.~\ref{Fig:1}). This rich excitonic landscape, together with the emergence of a direct band gap in the monolayer limit~\cite{Manzeli2017,Jing2020}, has stimulated intense research activity over the past decade, with important implications for both fundamental condensed-matter physics and technological applications~\cite{Gong2017,Mak2016,Ciarrocchi2022,Schaibley2016,Mueller2018}.

Beyond their central role in optical absorption and emission, excitons in TMDs underpin a variety of emerging concepts such as excitonics, in which excitons are exploited as information carriers, and valleytronics, where the valley degree of freedom associated with inequivalent K and K$'$ points in the Brillouin zone enables novel paradigms for information processing~\cite{Liu2019}. Recent advances in material synthesis have enabled the fabrication of high-quality, large-area [millimeter-scale, Fig.~\ref{Fig:1}(b)] TMD crystals, as well as the realization of van der Waals heterostructures in which controlled stacking and twisting of single-crystal layers [Fig.~\ref{Fig:1}(a)]~\cite{Liu2020} make it possible to induce correlated and optoelectronic phenomena characteristic of the emerging fields of twistronics and Moir\'e engineering~\cite{Carr2020,Regan2022,Li2024}. These developments have motivated extensive investigations of exciton formation, relaxation, and transport dynamics using ultrafast spectroscopic techniques, revealing a rich interplay between charge carriers, excitons, and lattice degrees of freedom on femtosecond timescales~\cite{Li2019,Trovatello2020,Selig2019,Waldecker2017}.

While these studies have established a detailed understanding of excitonic physics in the perturbative regime, extending this knowledge to the strong-field and sub-cycle domains remains an open challenge. This currently limits potential applications in the emerging fields of field-driven and petahertz electronics, which aim to exploit coherent light--matter interactions on attosecond timescales, where the electronic response follows the instantaneous electric field of light~\cite{Heide2024,Cavaletto2025}. Bridging the gap between the femtosecond excitonic dynamics extensively studied in TMDs and the attosecond regime relevant to field-driven electronics would therefore open new opportunities for ultrafast control of quantum degrees of freedom, including excitonic and valley pseudospins~\cite{Gucci2026,Slobodeniuk2023,Kim2014,Quintela2025,Kamath2026}.

Attosecond pump--probe techniques based on extreme-ultraviolet (XUV) and soft x-ray (SXR) radiation provide direct access to such sub-cycle dynamics~\cite{Borrego2022,Zong2023,Agostini2024,Lhuillier2024,Krausz2024,Alexander2025}. However, their application to low-dimensional materials remains limited. Only a few examples of attosecond pump--probe spectroscopy applied to TMDs can be found in the literature~\cite{Attar2020,Chang2021,Buades2021,Schumacher2023,Oh2023}, all of which were performed on relatively thick films (on the order of 100\,nm, i.e., bulk-like samples). Moreover, these investigations primarily focused on the fs-to-ps response following photoexcitation, without addressing the coherent phenomena occurring during the interaction with a few-cycle driving pulse. In addition to the experimental challenges associated with reduced sample thickness, a major obstacle is the limited knowledge of the static optical properties of these materials in the XUV and SXR spectral ranges. Such information is essential for the correct design and interpretation of attosecond experiments, including transient absorption and reflection spectroscopy~\cite{Geneaux2019,DiPalo2024,Inzani2025}. Consequently, despite the extensive characterization of TMDs in the visible and IR regimes, their optical response at high photon energies remains largely unexplored, particularly in the monolayer limit, hindering the application of attosecond techniques to investigate their sub-cycle dynamics.

An additional open question concerns the nature of electronic excitations in this spectral range. While excitons dominate the optical response of TMDs at low energies, it remains unclear whether analogous core-excitonic resonances persist at XUV photon energies, where optical transitions involve shallow core levels and unoccupied conduction-band states. Addressing this issue is crucial to determine whether concepts developed in the visible regime, such as excitonics and valley-selective light--matter interactions, can be extended to the XUV and attosecond domains~\cite{Moulet2017,Geneaux2020,Lucchini2021,Geondzhian2022,Gannan2025}.

In this work, we contribute to addressing these open questions by reporting, to the best of our knowledge, the first experimental determination of the complex refractive index of monolayer (1L) and bilayer (2L) \mos in the XUV spectral range. Our results provide a quantitative description of the optical response around the Mo N$_{2,3}$ edge, enabling accurate modeling of light--matter interactions in this energy range. Furthermore, we show that, in contrast to the visible and IR spectral ranges, the XUV response of \mos does not exhibit pronounced excitonic resonances. Instead, the absorption spectrum is characterized by a broad onset associated with dipole-allowed transitions from Mo $4p$ core levels to conduction-band states.

\begin{figure*}[htbp]
	\begin{center}
		\begin{tabular}{c}
			\includegraphics[width=0.95\textwidth]{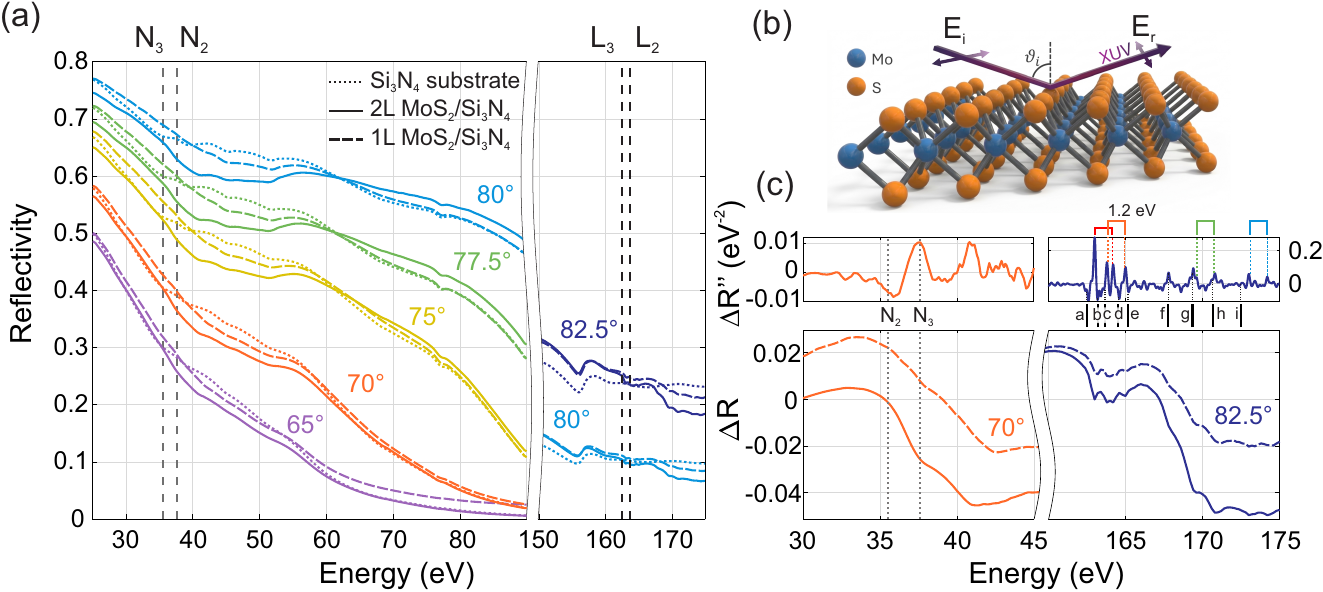}
		\end{tabular}
	\end{center}
	\caption[example]{\label{Fig:2} 
		Multi-angle reflectivity in the XUV and SXR region. \textbf{(a)} Reflectivity spectra of bulk \sn substrate (dotted curves), 1L (dashed curves) and 2L \mos (solid curves) deposited on \sn substrate for different incidence angles, $\theta_i$, with respect to the sample surface normal. Vertical dashed grey lines indicate the position of N$_2$ (37.6\,eV) and N$_3$ (35.5\,eV) absorption edges of Mo \cite{Kortright2001}. The black dashed vertical lines indicate instead the L$_2$ (163.6\,eV) and L$_3$ (162.5\,eV) absorption edges of S \cite{Thompson2009}. \textbf{(b)} Cartoon of the reflection geometry showing the definition of $\theta_i$ and the relative orientation of the crystal layers with respect to the radiation electric fields. \textbf{(c)} Main panel, reflectivity difference (target minus bare substrate), $\Delta R$, of 1L- (dashed) and 2L-\mos (solid) samples for two selected values of $\theta_i$. The kinks of both curves relate to the Mo N$_{2,3}$ and S L$_{2,3}$ edges as highlighted by $\Delta R''$, corresponding to the second derivative of the bilayer $\Delta R$, which is plotted in the upper panels. The S L$_{2,3}$ edge has been further subdivided. The black markers labeled from a to i report the transitions identified in Ref.~\cite{Parija2018}, while the colored dotted lines indicate a series of extrema that we found to be separated by the spin-orbit split coupling constant of sulfur, i.e. 1.2\,eV \cite{Han2011}.  
	}
\end{figure*}

We further compare the experimental results with theoretical calculations performed at different levels of approximation. A preview of the performance of these methods in the optical spectral range is shown in Fig.~\ref{Fig:1}(c), and will be discussed in more detail later in the text. The analysis in the XUV regime reveals that: (i) simplified models based on a limited number of bands and inadequate treatment of local-field screening predict artificially strong peaks, not supported by the experimental evidence, that may lead to an overestimation of the contribution of excitonic transitions; and (ii) local-field effects play a major role in suppressing exciton formation in the XUV regime of \mos, thereby explaining the absence of sharp core-excitonic resonances. Comparison with the existing literature further suggests that this behavior may be common among Mo-based TMDs, providing important constraints for theoretical modeling and for the interpretation of attosecond spectroscopies in these compounds.

\section{Experimental methods}

Both 1L and 2L \mos crystals investigated in this work were fabricated following the procedure reported in Ref.~\cite{Liu2020} and deposited onto bulk amorphous \sn substrates. To retrieve their complex optical constants, we performed multi-angle reflectivity measurements at the BEAR beamline of the Elettra synchrotron radiation facility~\cite{Nannarone2004}. The experimental setup is equipped with a high-resolution monochromator that enables the selection of photon energies spanning the entire XUV-to-SXR spectral range, from 3 to 1500\,eV, together with a reflectometer allowing precise measurements of absolute reflectivity at arbitrary incidence angles $\theta_i$~\cite{Pasquali2004}.

The intensities of the incident ($I_0$) and reflected ($I_R$) radiation were measured using detectors with calibrated readout currents directly proportional to the incident photon flux. After subtraction of the detector background signal (dark current), the ratio $R = I_R/I_0$ yields the absolute reflectivity of the sample. For all measurements reported here, the synchrotron radiation was $s$-polarized, corresponding to an electric field oscillating parallel to the plane of the \mos layers [see Fig.~\ref{Fig:2}(b)].

Reflectivity spectra were acquired with an energy step of 0.1\,eV, with an estimated uncertainty below 2\% (relative error) in the worst-case scenario. Different combinations of gratings and filters were employed to optimize radiation selection in different photon-energy regions: Sn filters for 20--25\,eV, Al for 40--75\,eV, Si for 70--100\,eV, and B filters for the 100--190\,eV range. The tabulated positions of the absorption edges of the different filters were used to calibrate the energy axis. Measurements acquired at the boundaries between different configurations were connected by averaging over a common energy interval.

The beam spot size on the sample was of the order of 100\,$\mu$m. Combined with the large lateral size of our samples [of the order of mm, see Fig.~\ref{Fig:1}(b)], this enabled reflectivity measurements close to grazing incidence without the need for extreme tight focusing.

\section{Reflectivity data}

Figure~\ref{Fig:2}(a) displays the reflectivity spectra measured in the 25--90\,eV and 150--175\,eV energy ranges, recorded as described in the previous section. Different colors indicate measurements acquired at different incidence angles $\theta_i$, defined with respect to the surface normal [see Fig.~\ref{Fig:2}(b)]. The dotted curves show the absolute reflectivity of the bare \sn substrate, while the results for 2L and 1L \mos are represented by solid and dashed curves, respectively.

Below 90\,eV, the reflectivity of the \sn substrate exhibits a smooth and featureless spectral dependence for all investigated values of $\theta_i$, consistent with the absence of pronounced absorption edges in this energy range. Both \mos samples display clear deviations from the substrate response, particularly between 30 and 60\,eV, indicating the interaction of the incident radiation with the atomically-thin \mos layers. This spectral region includes the Mo N$_3$ and N$_2$ absorption edges, located at photon energies of 35.5 and 37.6\,eV, respectively~\cite{Kortright2001}. These features originate from dipole-allowed transitions induced by XUV photon absorption, promoting electrons from the spin-orbit-split Mo $4p_{3/2}$ and $4p_{1/2}$ semicore levels (separated by 2.1\,eV) into unoccupied conduction-band states~\cite{Fuggle1980,Kumar2020}.

To further substantiate the interaction of light with the \mos atomic layers, reflectivity measurements were extended into the SXR spectral region, covering photon energies between 150 and 175\,eV, which include the sulfur L$_3$ and L$_2$ absorption edges at 162.5 and 163.6\,eV, respectively~\cite{Thompson2009}. At these higher photon energies, the absolute reflectivity is significantly reduced because of the stronger interaction between SXR radiation and matter, restricting meaningful measurements to incidence angles above 80$^\circ$. The corresponding reflectivity spectra, acquired at $\theta_i = 80^\circ$ and $82.5^\circ$, are shown in Fig.~\ref{Fig:2}(a). Both \mos samples exhibit distinct spectral features in the vicinity of the sulfur L$_2$ and L$_3$ edges, in stark contrast to the nearly flat reflectivity of the bare \sn substrate.

The contribution of the \mos layers to the overall reflectivity becomes particularly evident when considering the differential reflectivity shown in Fig.~\ref{Fig:2}(c) for selected values of $\theta_i$, defined as $\Delta R = R_{\ce{MoS2}} - R_{\ce{Si3N4}}$, where $R_{\ce{MoS2}}$ and $R_{\ce{Si3N4}}$ denote the reflectivity of the \mos-covered sample and of the bare substrate, respectively. This representation isolates the optical response of the \mos layers and reveals a systematically smaller signal (in absolute value) for the 1L-\mos sample (dashed curves) compared with the 2L-\mos sample (solid curves).

Furthermore, $\Delta R$ exhibits clear kinks in the vicinity of the absorption edges, which are more clearly highlighted by the local extrema of its second derivative with respect to energy, $\Delta R'' = \partial^2 \Delta R / \partial \omega^2 $, reported in the upper panel of Fig.~\ref{Fig:2}(c). Below 45\,eV, two pronounced extrema coincide with the Mo N$_2$ and N$_3$ edges (vertical gray dashed lines). In addition, a clear peak in $\Delta R''$ is observed at approximately 40.9\,eV, which could be associated with transitions to higher-energy conduction-band states~\cite{Trainer2017}.

In the SXR range (S L-edge), $\Delta R''$ for both the 1L and 2L samples reveals several negative and positive peaks (black markers labeled from \textit{a} to \textit{i}) that coincide with dipole-allowed transitions from S $2p$ states to unoccupied electronic states near the bottom of the conduction band, as previously identified by Parija \textit{et al.} in 45-nm-thick \mos films~\cite{Parija2018}. Interestingly, we also observe additional extrema, marked by colored dotted lines, forming a series of pairs separated by approximately 1.2\,eV, corresponding to the spin-orbit splitting of the sulfur $2p_{1/2}$ and $2p_{3/2}$ core levels~\cite{Han2011}. Similarly to the N$_2$ and N$_3$ edges, these features may be associated with transitions to higher-energy conduction-band states.

\section{Data analysis and discussion}

\begin{figure}[htbp]
	\begin{center}
		\begin{tabular}{c}
			\includegraphics[width=0.47\textwidth]{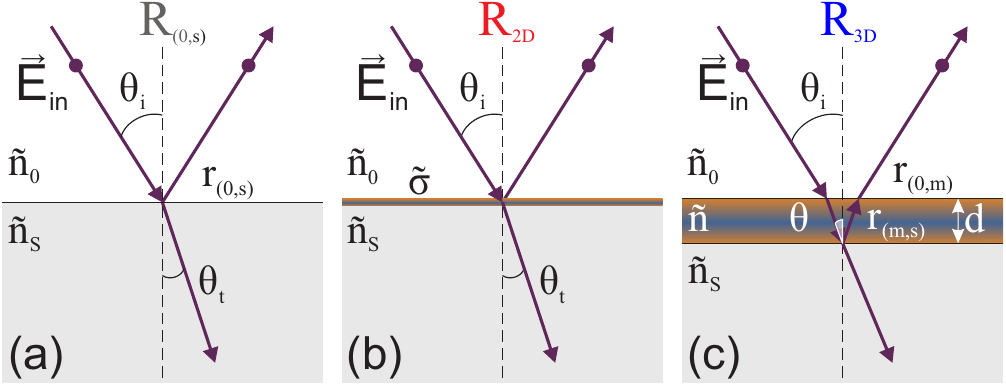}
		\end{tabular}
	\end{center}
	\caption[example]{\label{Fig:c} 
		Schematic representation of the reflection happening on the \sn substrate \textbf{(a)}, and on the \mos samples modeled with a 2D single-sheet of surface conductivity $\tilde{\sigma}$, \textbf{(b)}, or with a 3D layer of thickness $d$, \textbf{(c)}.   
	}
\end{figure}

While reflection from a homogeneous surface defined by a bulk sample can be accurately described by the Fresnel equations down to atomic and nanometric length scales~\cite{Lucchini2015} [Fig.~\ref{Fig:c}(a)], two-dimensional and layered materials require a different theoretical treatment~\cite{Yoo2022}.

If the radiation wavelength $\lambda$ is much larger than the product of the real refractive index $n$ and the geometrical thickness $d$ of the 2D layer, i.e., $\lambda \gg nd$, the thin layer can be modeled as a two-dimensional conductive sheet with complex conductivity $\tilde{\sigma}$ [Fig.~\ref{Fig:c}(b)]. By contrast, when the material consists of several atomic layers, the sample is often described as a three-dimensional slab of finite thickness $d$ [Fig.~\ref{Fig:c}(c)]~\cite{Pasquali2014,Capelli2016,Pasquali2021}.

Interestingly, when the optical phase accumulated by light during propagation through the thin layer is much smaller than unity, the two descriptions yield consistent results provided that the anisotropy of the material is properly taken into account~\cite{Majrus2018}. In the following, we first characterize the optical properties of the \sn substrate using the model depicted in Fig.~\ref{Fig:c}(a). We then use these results to apply the single-sheet model shown in Fig.~\ref{Fig:c}(b) and extract the optical constants of the \mos layer. Finally, we employ the 3D slab model of Fig.~\ref{Fig:c}(c) as an internal consistency check and to validate our description of 1L \mos.

\subsection{Substrate characterization}

\begin{figure}[htbp]
	\begin{center}
		\begin{tabular}{c}
			\includegraphics[width=0.41\textwidth]{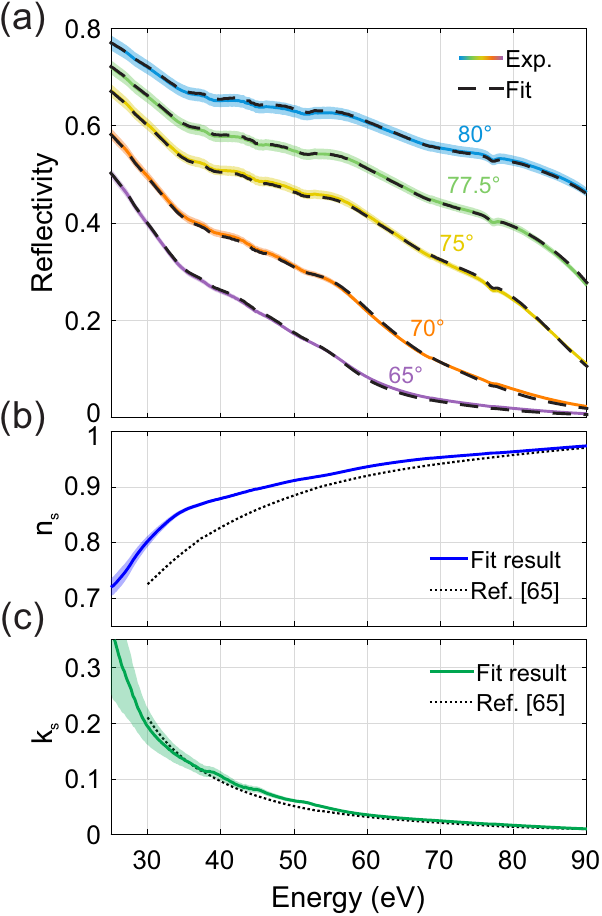}
		\end{tabular}
	\end{center}
	\caption[example]{\label{Fig:3} 
		\sn substrate characterization in the XUV region. \textbf{(a)} Fit results of the experimental multi-angle reflectivity data for the bulk \sn substrate  (solid colored curves) with the Fresnel model of Eq.~\eqref{Eq:Rsub} (black dashed curves). The shaded areas around the experimental curves correspond to the estimated 2\% uncertainty. \textbf{(b)}, Real (blue solid) and, \textbf{(c)}, imaginary (green solid) part of the complex substrate refractive index $\tilde{n_s} = n_s+ik_s$ as retrieved by the non-linear fitting procedure. In both panels the black dotted curves correspond to the tabulated values reported in Ref.~\cite{Henke1993}. The shaded areas indicate the accuracy of the retrieval (see main text).
	}
\end{figure}

Using the well-known Snell equation and Fresnel formula for $s$-polarized light, the vacuum/substrate reflectivity $R_{\left(0,\mathrm{s}\right)}$ can be written as:
\begin{eqnarray}
	R_{\left(0,\mathrm{s}\right)}(\omega) &=&|r_{\left(0,\mathrm{s}\right)}(\omega)|^2 =\nonumber \\
	&=&\left|\frac{\tilde{n}_0\cos(\theta_i)-\sqrt{\tilde{n}_{\mathrm{s}}(\omega)^2-\tilde{n}_0^2\sin(\theta_i)^2} }{\tilde{n}_0\cos(\theta_i)+\sqrt{\tilde{n}_{\mathrm{s}}(\omega)^2-\tilde{n}_0^2\sin(\theta_i)^2}}\right|^2,\label{Eq:Rsub}
\end{eqnarray}
where $\hbar\omega$ is the photon energy, $r_{\left(0,\mathrm{s}\right)}$ denotes the Fresnel reflection coefficient at the vacuum/\sn interface, and $\tilde{n}_0$ and $\tilde{n}_{\mathrm{s}}$ are the complex refractive indices of vacuum and substrate, respectively. While $\tilde{n}_0$ can be assumed equal to 1, $\tilde{n}_{\mathrm{s}}$ can be retrieved from the multi-angle reflectivity data using a nonlinear fitting procedure based on Eq.~\eqref{Eq:Rsub}~\cite{Kaplan2019}. The results are reported in Fig.~\ref{Fig:3}(a).

At each photon energy, the fitting procedure independently retrieves the real and imaginary components of $\tilde{n}_{\mathrm{s}} = n_{\mathrm{s}}+ik_{\mathrm{s}}$ by minimizing the deviation between the experimental reflectivity [solid curves, identical to those shown in Fig.~\ref{Fig:2}(a)] and the modeled reflectivity (black dashed curves) over all measured values of $\theta_i$. The retrieved values of $n_{\mathrm{s}}$ and $k_{\mathrm{s}}$ are shown in Figs.~\ref{Fig:3}(b) and \ref{Fig:3}(c) as colored solid curves, respectively, and exhibit a smooth spectral behavior consistent with the tabulated values for \sn (black dotted curves)~\cite{Henke1993}.

To quantify the associated uncertainty, we follow the procedure introduced by Kaplan \textit{et al.}~\cite{Kaplan2019}. Starting from the set of five reflectivity spectra $\left\{R_1,R_2,\dots{},R_5\right\}$ measured at incidence angles $\left\{\theta_1,\theta_2,\dots{},\theta_5\right\}$, we generate $3^5=243$ synthetic datasets by perturbing each spectrum according to $R_i(E)\pm \mathit{\delta R}(E)$, where $\mathit{\delta R}$ corresponds to the 2\% relative experimental uncertainty. The fitting procedure is then applied independently to each synthetic dataset, yielding a distribution of complex refractive index values. The solid curves in Figs.~\ref{Fig:3}(b) and \ref{Fig:3}(c) represent the resulting mean values, while the shaded areas denote the associated standard deviations, thereby providing a robust estimate of the confidence intervals of the extracted optical response.

\subsection{\mos optical constants}
\label{Sec:optconst}

Modeling atomically thin samples as a 2D single-sheet layer with complex conductivity $\tilde{\sigma}(E)$ is an approach that has been validated both experimentally~\cite{Nair2008} and theoretically~\cite{Merano2016}, primarily in the visible and IR spectral regions. Within this framework, the reflectivity of a system composed of a bulk substrate coated with a 2D surface layer can be expressed as:
\begin{equation}
	\label{Eq:R2D}
	R_{\mathrm{2D}}=\left|\frac{\tilde{n}_0 \cos(\theta_i)-\sqrt{\tilde{n}_{\mathrm{s}}^2-\tilde{n}_{0}^2\sin(\theta_i)^2}-\eta\tilde{\sigma}}{\tilde{n}_0 \cos(\theta_i)-\sqrt{\tilde{n}_{\mathrm{s}}^2-\tilde{n}_{0}^2\sin(\theta_i)^2}+\eta\tilde{\sigma}}\right|^2
\end{equation}
where $\eta$ is the vacuum impedance, related to the vacuum permittivity $\varepsilon_0$ and the speed of light $c$ through $\eta=1/(\varepsilon_0\,c)$. In this formulation, the presence of the 2D layer introduces a correction to the bare substrate reflectivity proportional to the complex sheet conductivity $\tilde{\sigma}=\sigma_1+i\sigma_2$. This contribution becomes particularly relevant in spectral regions where light--matter interaction is strong, such as in the vicinity of core-level absorption edges. Nevertheless, the phase accumulated during propagation is expected to remain $\ll 1$ across the entire XUV energy range for both \mos samples, thereby justifying the adoption of the 2D single-sheet model.
\begin{figure}[b]
	\begin{center}
		\begin{tabular}{c}
			\includegraphics[width=0.47\textwidth]{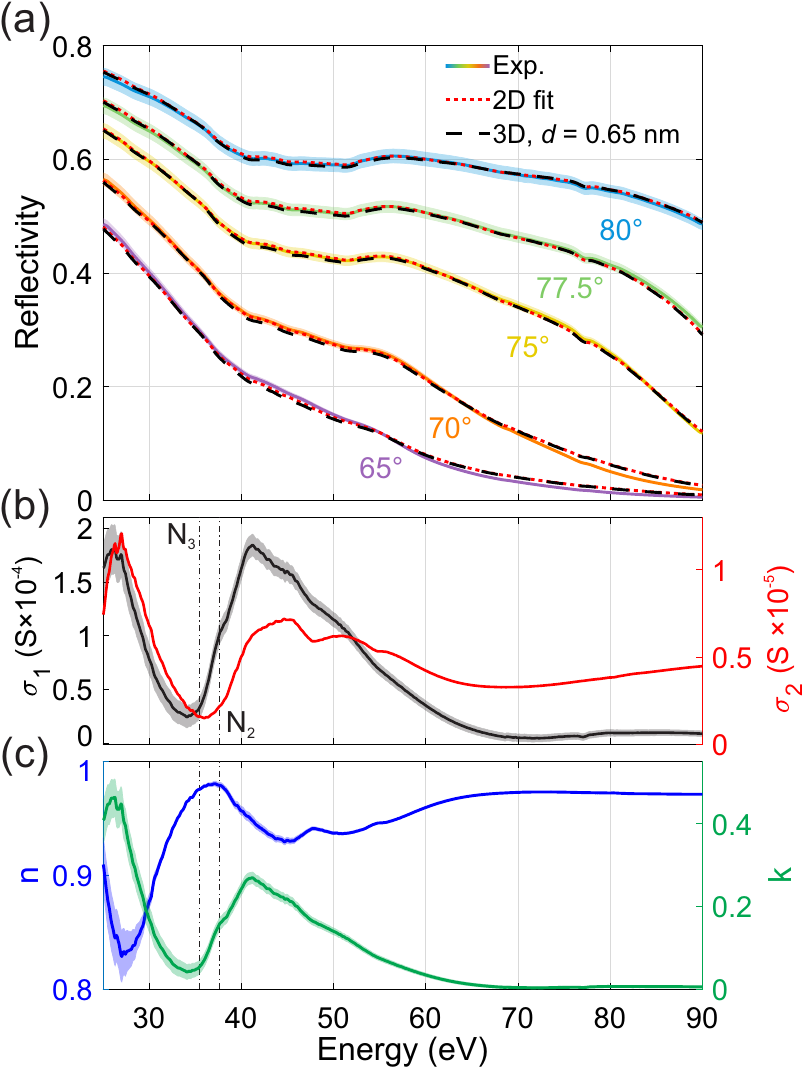}
		\end{tabular}
	\end{center}
	\caption[example]{\label{Fig:4} 
		Complex optical response of 2L-\mos in the XUV region. \textbf{(a)} Fit results of the experimental multi-angle reflectivity data (solid colored curves) with the 2D single-sheet model (red dotted curves). Black-dashed curves represent the reflectivity spectra calculated with the 3D film model starting from the 2D single-sheet fit results and assuming a crystal thickness $d =  0.650$\,nm. As in Fig.~\ref{Fig:3}(a), the shaded areas around the experimental curves correspond to the estimated 2\% uncertainty. \textbf{(b)} Real (black) and imaginary (red) part of the optical conductivity, $\tilde{\sigma} = \sigma_1+i\sigma_2$, extracted from the fit based on the 2D model, along with shaded areas indicating accuracy of the retrieval (see text). \textbf{(c)} Real (blue) and imaginary (green) part of the effective refractive index $\tilde{n} = n+ik$, obtained from the optical conductivity of panel (b) when assuming a layer thickness of $d =  0.650$\,nm. In both panels the dash-dotted vertical lines indicate the Mo N$_2$ and N$_3$ edges. 
	}
\end{figure}

We note that, since the complex conductivity and polarizability $\tilde{\alpha}$ are related through
\begin{equation}
	\tilde{\alpha}(\omega) = \frac{i}{\omega}\tilde{\sigma}(\omega),\label{Eq:sig_alp}
\end{equation}
any resonance observed in $\sigma_1$ reflects the absorptive component of the optical response, \ima, which is also encoded in the imaginary part of the refractive index, $k$.

Starting from the complex substrate refractive index $\tilde{n}_{\mathrm{s}}$ obtained in Figs.~\ref{Fig:3}(b) and \ref{Fig:3}(c), we fitted the experimental reflectivity spectra of the 2L samples [colored curves in Fig.~\ref{Fig:4}(a)] using Eq.~\eqref{Eq:R2D} and the same multi-angle fitting procedure introduced above. The fitting results, shown as red dotted curves in Fig.~\ref{Fig:4}(a), reproduce the experimental reflectivity spectra (colored solid curves) within the 2\% experimental uncertainty (shaded areas) for all measured incidence angles. The retrieved complex conductivity is reported in Fig.~\ref{Fig:4}(b), where the shaded areas representing the confidence interval were calculated using the same procedure adopted for the substrate characterization (see previous section).

Having obtained a complex-valued optical conductivity $\tilde{\sigma}$ and polarizability $\tilde{\alpha}$ that consistently reproduce the reflectivity data of the 2L-\mos sample, we next sought to express the optical response in terms of an effective relative dielectric function and complex refractive index, $\tilde{n}=\sqrt{\varepsilon_r(\omega)} = n+ik$, which constitute a more conventional description in optical spectroscopy. The effective relative dielectric function is defined as
\begin{equation}
	\varepsilon_r(\omega) = 1+\frac{\tilde{\alpha}(\omega)}{\varepsilon_0d},\label{Eq:sigma_n}
\end{equation}
where $d$ represents the effective thickness of the material~\cite{Li2018}. While the concept of thickness is unambiguous for bulk crystals, its definition in atomically thin systems is less straightforward and is commonly associated with the interlayer spacing in few-layer materials~\cite{Majrus2018}. For bilayer \mos, reported values of the interlayer spacing range between 0.6 and 0.7\,nm, as determined by both experimental~\cite{Wang2023} and theoretical~\cite{Xiao2014} studies.

Figure~\ref{Fig:4}(c) shows the real (blue) and imaginary (green) parts of the effective refractive index of 2L \mos, calculated from Eq.~\eqref{Eq:sigma_n} using the retrieved $\tilde{\sigma}$ of Fig.~\ref{Fig:4}(b) and assuming an effective thickness of $d = 0.650$\,nm. To verify the consistency of these assumptions, we used the resulting $\tilde{n}$ to calculate the reflectivity spectra within a 3D slab model~\cite{Parratt1954} and directly compare them with the experimental data.

Within this framework, illustrated in Fig.~\ref{Fig:c}(c), light undergoes partial reflection at both the vacuum/\mos and \mos/\sn interfaces, with reflection coefficients $r_{\left(0,\mathrm{m}\right)}$ and $r_{\left(\mathrm{m},\mathrm{s}\right)}$, respectively. In addition, during the double passage through the \mos layer, the field accumulates a phase $2\delta$ that depends on the propagation angle inside the layer, $\theta$:
\begin{equation}
	2\delta = \frac{4\pi d}{\lambda}\,\tilde{n}\cos(\theta). \label{Eq:delta}
\end{equation}
The total reflectivity of the system can then be written as:
\begin{equation}
	R_{\mathrm{3D}} = \left|\frac{r_{\left(0,\mathrm{m}\right)}+r_{\left(\mathrm{m},\mathrm{s}\right)}e^{i2\delta}}{1+r_{\left(0,\mathrm{m}\right)}r_{\left(\mathrm{m},\mathrm{s}\right)}e^{i2\delta}}\right|^2. \label{Eq:R3D}
\end{equation}

The reflectivity spectra calculated with Eq.~\eqref{Eq:R3D}, using the complex refractive index shown in Fig.~\ref{Fig:4}(c), are reported in Fig.~\ref{Fig:4}(a) as black dashed curves. As can be observed, the two models are in close quantitative agreement, thereby justifying the assumed effective thickness of $d = 0.650$\,nm and confirming the consistency between the extracted complex conductivity $\tilde{\sigma}$ and the corresponding complex refractive index $\tilde{n}$.

We note that, although the 2D and 3D models are mutually consistent, directly retrieving the complex refractive index $\tilde{n}(\omega)$ through a multi-angle fit of the experimental reflectivity spectra using Eq.~\eqref{Eq:R3D} leads to significantly poorer convergence. This result further underscores the robustness and suitability of the 2D complex-conductivity approach for atomically thin systems in the XUV regime.

We next attempted to extend the fitting procedure to the 1L-\mos sample. Unfortunately, owing to the reduced contribution of a single atomic layer to the total reflectivity, the monolayer signal is intrinsically weaker than that of the bilayer, resulting in poorer convergence of the fitting algorithm when $\tilde{\sigma}(\omega)$ is treated as a free parameter. We therefore adopted a scaling approach based on the optical conductivity extracted from the 2L-\mos dataset using the 2D sheet model [Fig.~\ref{Fig:4}(b)].

Although the optical properties of \mos in the IR-visible spectral range are strongly thickness dependent because of the transition from a direct to an indirect band gap~\cite{Mak2010}, the response in the XUV spectral region is dominated by atomic-like transitions originating from (semi-)core electronic states, specifically the Mo $4p$ levels in the present case~\cite{Fuggle1980}. At these photon energies, the optical response is expected to scale approximately linearly with the number of atoms participating in the interaction~\cite{Henke1993}, thereby justifying the use of a rescaled 2L conductivity to model the 1L response.

Accordingly, the reflectivity of the monolayer sample can be estimated using the same formalism introduced in Eq.~\eqref{Eq:R2D}, together with the substitution
\begin{equation}
	\tilde{\sigma}_{2L}(\omega)\rightarrow \tilde{\sigma}_{1L}(\omega)\equiv\gamma \tilde{\sigma}_{2L}(\omega),\label{Eq:2L_1L}
\end{equation}
where $\gamma$ is a scaling factor.

\begin{figure}[htbp]
	\begin{center}
		\begin{tabular}{c}
			\includegraphics[width=0.45\textwidth]{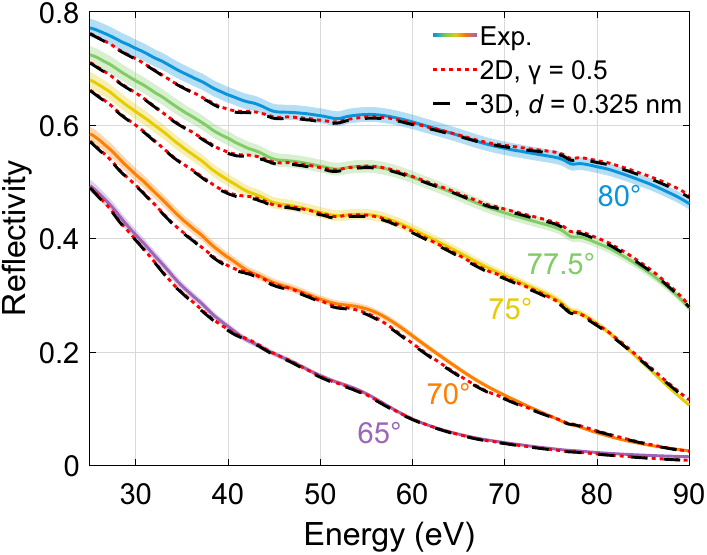}
		\end{tabular}
	\end{center}
	\caption[example]{\label{Fig:5} 
		Experimental reflectivity spectra of the 1L-\mos sample [colored solid curves, same as in Fig.~\ref{Fig:2}(a)] together with the reflectivity simulated with the 2D single-sheet model while considering half of the complex conductivity found with the 2L-\mos sample (red dotted curves), and with the 3D slab model starting from the $n$ and $k$ of Fig.~\ref{Fig:4}(c) and assuming an effective thickness $d = 0.325$\,nm (black dotted curves). The good agreement between experiment and simulations highlights the validity of the complex optical constants reported in Fig.~\ref{Fig:4}. 
	}
\end{figure}

Figure~\ref{Fig:5} shows the experimental reflectivity spectra of the 1L-\mos sample (colored curves with shaded areas) together with the spectra calculated using the 2D model [Eq.~\eqref{Eq:R2D}] assuming a scaling factor $\gamma = 0.5$ (red dotted curves), i.e., by considering half of the complex conductivity extracted from the 2L sample. The calculated spectra reproduce the experimental data within the 2\% uncertainty. This agreement is particularly remarkable considering that the measurement is sensitive to the homogeneity of the illuminated sample area, which extends over several millimeters at grazing incidence.

The black dashed curves instead represent the reflectivity spectra calculated using the 3D model of Eq.~\eqref{Eq:R3D}, under the assumption of both halved effective thickness ($d = 0.325$\,nm) and halved complex conductivity ($0.5\tilde{\sigma}_{2L}$). Also in this case, the two models are in close quantitative agreement.

We finally note that, although our analysis shows that the XUV response of the 1L-\mos sample is, within the experimental uncertainty, well described by a complex conductivity equal to half that of the 2L-\mos sample, the two systems share the same effective complex refractive index. Indeed, the result of Eq.~\eqref{Eq:sigma_n} remains unchanged when both $d$ and $\tilde{\sigma}$ are scaled by the same factor.

\subsection{Discussion}

We now turn to the physical interpretation of the complex optical response of \mos extracted from the reflectivity analysis. Both the real part of the optical conductivity, $\sigma_1(\omega)$ [black curve in Fig.~\ref{Fig:4}(b)], and the imaginary part of the complex refractive index, $k(\omega)$ [green curve in Fig.~\ref{Fig:4}(c)], exhibit a broad resonance whose onset coincides with photon energies near the Mo N$_2$ and N$_3$ absorption edges. This behavior provides a clear signature of photon absorption in this spectral region and indicates the presence of dipole-allowed electronic transitions from the spin-orbit-split Mo $4p_{1/2}$ and $4p_{3/2}$ semicore states into unoccupied electronic states at the bottom of the conduction band.

Density functional theory (DFT) calculations reported in the literature consistently show that, within several eV above the conduction-band minimum, the electronic density of states is dominated by contributions from Mo $4d$ and S $3p$ orbitals~\cite{Ulian2023,Yamusa2022}. We therefore conclude that the optical transitions responsible for the observed XUV absorption lineshape primarily involve Mo $4d$-derived conduction-band states, with the largest contributions arising from the d$_{z^2}$, d$_{x^2+y^2}$, and d$_{xy}$ orbitals.

It is worth noting that no sharp peaks are observed below the N edges. Indeed, in agreement with the analysis of $\Delta R''$, the spectrum shows a broad feature centered at about 40.9\,eV, with a shoulder close to the N$_2$ resonance. This suggests that excitonic transitions carry a significantly weaker spectral weight in the XUV optical response of 1L and 2L \mos compared with what is observed in the visible/IR spectral range. A shallow Mo N$_{2,3}$ edge without sharp resonances was also reported by T. J. Whitcher \textit{et al.}, who studied the optical properties of bulk \mos samples as a function of temperature in the spectral range from 0.6 to about 45\,eV~\cite{Whitcher2021}.

Our findings are also consistent with previous XUV absorption studies on Mo-based TMD thin films, suggesting that our conclusions are not limited to \mos, but may apply more broadly to this family of materials. For example, Attar \textit{et al.} reported the experimental absorbance of a 50-nm-thick 2H-\ce{MoTe2} film in a spectral region encompassing both the Mo and Te absorption edges (39--46\,eV)~\cite{Attar2020}. As in the present case, the absorption spectrum displays a pronounced onset at photon energies close to the Mo edge, followed by a broad spectral feature extending toward higher energies before merging with contributions from the chalcogen absorption edges.

A closely related behavior was also observed in the work of Schumacher \textit{et al.} on attosecond XUV absorption spectroscopy of \ce{MoSe2} thin films~\cite{Schumacher2023}. Although the static absorption spectrum is not explicitly shown, the reported time-resolved response at the Mo edge closely resembles the spectral features observed in Ref.~\cite{Attar2020}, suggesting a comparable XUV optical response across different Mo-based TMDs.

In contrast, a qualitatively different behavior has been reported for W-based TMDs. Chang \textit{et al.} performed time-resolved XUV absorption spectroscopy on a 40-nm-thick \ce{WS2} film in the vicinity of the W N$_{6,7}$ (32--37\,eV) and O$_3$ (37--45\,eV) absorption edges~\cite{Chang2021}. In addition to a broad absorption feature starting at approximately 38.5\,eV, attributed to transitions from the W 5$p_{3/2}$ level, the authors observed several sharp Fano-like resonances at lower photon energies. These features were assigned to transitions from the W $4f_{7/2}$ and $4f_{5/2}$ semicore states into the conduction band and interpreted as signatures of semicore excitons generated by XUV photon absorption.

Notably, these excitonic resonances exhibit a markedly different dynamical behavior in pump-probe experiments compared with the continuum-like response observed near the absorption edge. Similar findings were later reported by the same group for \ce{WSe2} thin films, further corroborating the emergence of excitonic effects in the XUV response of W-based TMDs~\cite{Oh2023}.

These observations highlight a consistent XUV optical response among Mo-based TMDs, persisting from thick films down to the few-layer and atomically thin limits investigated here, while revealing a qualitatively distinct behavior in W-based compounds due to the emergence of strongly bound semicore excitons.

\section{Numerical Methods}

To gain deeper insight into the microscopic mechanisms governing the macroscopic optical response of \mos in this photon-energy range, we complemented our experimental analysis with theoretical simulations. The optical properties of \mos have been extensively investigated theoretically in the visible/UV spectral range~\cite{MolinaSanchez2013,Qiu2013,Mak2010}. In the monolayer limit, the optical absorption in the visible region is dominated by two features, commonly referred to as the A and B peaks, located well below the band gap and governed by electron-hole (e-h) interactions, together with a broader high-energy feature commonly labeled C [see also Fig.~\ref{Fig:1}(c)]. Their theoretical description is based on the solution of the Bethe-Salpeter equation (BSE), which accounts for e-h correlations through an eh-exchange term and a screened eh-direct term. The latter is responsible for the formation of bound excitonic states. Once the BSE is solved, the polarizability $\tilde{\alpha}(\omega)$ (or, equivalently, the dielectric function in 3D materials) can be directly constructed~\cite{Onida2002}.

Alternatively, optical properties can be described by propagating a time-dependent (TD) equation of motion for the density matrix, in which the probe pulse is explicitly included in the simulations. 
In this approach, the polarizability of the material, $\tilde{\alpha}(\omega)$, is reconstructed from the time-dependent polarization (or current) induced by the probe laser pulse. Within the TD formalism, the eh-exchange interaction and the screened eh-direct interaction are captured through the update of the Hartree (H) potential and the screened exchange (SEX) self-energy, respectively. Accordingly, the scheme is commonly referred to as TD-HSEX~\cite{Attaccalite2011} or, alternatively, adiabatic-GW~\cite{Chane2021}, since SEX corresponds to the adiabatic part of the well-known GW self-energy~\cite{Onida2002}.
BSE and TD-HSEX yield identical results when the same approximations and the same number of bands are employed~\cite{Attaccalite2011}.

In the visible/IR spectral range, the A and B excitons can be accurately described by including only a few valence and conduction bands (CBs). BSE and TD-HSEX can be solved either fully \ai, on top of DFT simulations, or by relying on alternative approaches such as Wannier-basis models and effective 2D interactions. Both strategies provide an excellent quantitative description of the optical spectra~\cite{MolinaSanchez2013,Qiu2013,Cistaro2023}.

In the present work, we perform fully \ai, BSE calculations together with TD-HSEX simulations based on effective 2D interactions. The TD-HSEX simulations are carried out using the Electron Dynamics and Ultrafast Spectroscopy (EDUS) approach~\cite{Cistaro2023,Malakhov2024,Mosquera2024}.
EDUS solves the one-particle density matrix in real time and operates in the so-called Wannier gauge, in which the one-particle Hamiltonian is not diagonal, thereby significantly reducing the computational cost (see Sec.~\ref{Sec:EDUS}). BSE calculations are instead performed using the \lumen, fork~\cite{Sangalli2026} of the \textit{Yambo} code~\cite{Sangalli2019}, which implements \ai, many-body perturbation theory (see Sec.~\ref{Sec:Lumen}). After benchmarking both approaches in the visible/IR spectral range, we extended them to the XUV regime and used them to model transitions from the Mo 4$p_{1/2}$ and 4$p_{3/2}$ core states to conduction-band states, including spin-orbit coupling. The results are presented and discussed in Sec.~\ref{Sec:NumRes}.

\subsection{EDUS}
\label{Sec:EDUS}

In the simulations using EDUS, we propagate the one-particle density matrix, $\rho\left(\mathbf{r},\mathbf{r}'\right)$, projected onto the Wannier basis set, $\rho_{nm\mathbf{k}}$, where $\mathbf{k}$ represents the crystalline momentum. Both diagonal and off-diagonal terms are retained. This allows us to formulate the TD-HSEX equation of motion in the so-called Wannier gauge, which reduces the computational cost while also enabling an accurate description of light--matter interaction beyond the linear regime~\cite{Cistaro2023}. The TD-HSEX equation of motion reads:
\begin{equation}
	\label{eq:eom_rho_EDUS}
	\begin{split}
		i \dot{\rho}_{nm\mathbf{k}}(t) =\;& 
		\left[ \hat{h}^{\mathrm{eff}}(t), \rho(t) \right]_{nm\mathbf{k}} + \\
		&+ i\, \boldsymbol{\mathrm{E}}(t) \cdot \nabla_{\mathbf{k}} \rho_{nm\mathbf{k}}(t)
		- i\, \Gamma_{nm{\bf k}} \rho_{nm\mathbf{k}}(t)
	\end{split}
\end{equation}
with $\hat{h}^{\mathrm{eff}} = \hat{h}_0 + \hat{v}^{\mathrm{H-RK}} + \hat{\Sigma}^{\mathrm{X-RK}} + \boldsymbol{\mathrm{E}}(t)\cdot\boldsymbol{\xi}$. Here, $\hat{h}_0$ is the equilibrium electronic Hamiltonian calculated using DFT with a screened range-separated hybrid (SRSH) functional~\cite{Camarasa2023}, $\boldsymbol{\mathrm{E}}(t)$ is the electric field of the light pulse interacting with the sample, and $\boldsymbol{\xi}$ is the Berry connection. $\hat{v}^{\mathrm{H-RK}}$ and $\hat{\Sigma}^{\mathrm{X-RK}}$ are the Hartree and exchange (Fock) terms, respectively. $\Gamma_{nm{\bf k}}$ is an effective parameter accounting for both decay and decoherence processes. In our simulations, the dominant contribution arises from the N$_{2,3}$ core decay, for which we adopt a decay width of 300\,meV, corresponding to a lifetime of approximately 2.2\,fs.

The equation of motion is formulated and solved in reciprocal space by discretizing the \textbf{k}-space grid. The DFT calculations are performed using Quantum Espresso~\cite{Giannozzi2017}, followed by Wannierization using the Wannier90 code~\cite{Pizzi2020}. The evolving electron correlations are contained in the terms $\hat{v}^{\mathrm{H-RK}}$ and $\hat{\Sigma}^{\mathrm{X-RK}}$, which depend on the out-of-equilibrium part of the density matrix. In the Wannier gauge, these terms take the form

\begin{equation}
	\label{eq:hartree_exchange}
	\begin{aligned}
		v^{\mathrm{H-RK}}_{nn\mathbf{k}}[\Delta \rho]  =\;&
		\sum_{m,\mathbf{k}',\, \mathbf{G}}
		\mathrm{Re}\!\left[
		e^{i \mathbf{G}\cdot(\boldsymbol{t}_m-\boldsymbol{t}_n)}
		\right]
		\, V^{\mathrm{RK}}_{\mathbf{G}} \,
		\Delta \rho_{mm\mathbf{k}'}(t) \\
		\\
		\Sigma^{\mathrm{X-RK}}_{nm\mathbf{k}}[\Delta \rho] =\;&
		- \sum_{\mathbf{k}',\, \mathbf{G}}
		e^{i (\mathbf{k}'-\mathbf{k}+\mathbf{G})\cdot(\boldsymbol{t}_m-\boldsymbol{t}_n)} \\
		&\times V^{\mathrm{RK}}_{\mathbf{k}-\mathbf{k}'+\mathbf{G}}
		\, \Delta \rho_{nm\mathbf{k}'}(t)
	\end{aligned}
\end{equation}

Here, $\mathbf{G}$ corresponds to the reciprocal lattice vectors, and the RK suffix emphasizes the use of a Rytova-Keldysh (RK) potential, i.e., an effective screened interaction of the form
\begin{eqnarray} \label{eq:RKpot}
	V^{\mathrm{RK}}_{\mathbf{q}} = \frac{e^2}{\varepsilon_{0}(\varepsilon_t+\varepsilon_b)A} \frac{1}{ |{\bf q}|\left(r_{0} |{\bf q}|+1 \right)}.
\end{eqnarray}

It is worth noting that, despite the use of an effective interaction, the exchange (X) term $\hat{\Sigma}^{\mathrm{X-RK}}$ captures the same physics as the $\hat{\Sigma}^{\mathrm{SEX}}$ self-energy more commonly employed in the \emph{ab initio} community.

The effective potential of Eq.~\ref{eq:RKpot} is derived for 2D materials, in which the static dielectric permittivities of the top and bottom surrounding media are given by $\varepsilon_{t}$ and $\varepsilon_b$, respectively. Because of the discretization of reciprocal space, the normalization also requires inclusion of the system area $A$. The effective parameter $r_0$ allows the screening strength to be modified in both the Hartree and Fock terms. While we keep the screened interaction fixed at $r_0 = 41.5$\,\AA\ in the exchange term $\hat{\Sigma}^{\mathrm{X-RK}}$, which dominates the visible/IR response and is primarily responsible for the emergence of the A and B excitons [see Fig.~\ref{Fig:1}(c), dash-dotted curves], we vary the screening of the Hartree term by changing the value of the parameter $r_0$. Note that, in the limit $r_0=0$, the bare Coulomb interaction for a 2D system is recovered. For the XUV simulations, we consider two different cases for $\hat{v}^{\mathrm{H-RK}}$: a screened interaction with $r_0 = 41.5$\,\AA\ and a nearly unscreened interaction with $r'_0 = 0.01r_0 = 0.415$\,\AA. We do not set $r_0 = 0$ exactly in order to avoid excessively large Hartree contributions, which would make the time propagation numerically unstable.

\subsection{\lumen}
\label{Sec:Lumen}

The BSE is implemented in \lumen in the form of an eigenvalue problem,
\begin{equation}
	\hat{H}^{\mathrm{exc}} | A^{\lambda}\rangle = E_{\lambda} |A^{\lambda}\rangle
\end{equation}
where $\hat{H}^{\mathrm{exc}}$ is an effective two-particles excitonic Hamiltonian, with eigenvalues $E_{\lambda}$ (the excitonic energies). Expressed in transitions space, the eigenvectors $|A^\lambda\rangle$ take the form $A^{\lambda}_{cv\mathbf{k}}$, and
$\hat{H}^{\mathrm{exc}}$ matrix elements are~\cite{Sangalli2019}:

\begin{multline}
	\label{eq:exc}
	H^{\mathrm{exc}}_{\substack{cv\mathbf{k}\\c'v'\mathbf{k}'}} =
	\left(\epsilon_{c\mathbf{k}}-\epsilon_{v\mathbf{k}}\right)
	\, \delta_{c,c'} \delta_{v,v'} \delta\left(\mathbf{k}-\mathbf{k}'\right)+ \\
	+ \left(v_{\substack{cv\mathbf{k}\\c'v'\mathbf{k}'}}-W_{\substack{cv\mathbf{k}\\c'v'\mathbf{k}'}}(0)\right)
\end{multline}

where $\epsilon_{n\mathbf{k}}$ are the electronic energies obtained from KS+GW corrections, and the indices $c$ and $v$ run over conduction and valence states, respectively. The term $\epsilon_{c\mathbf{k}}-\epsilon_{v\mathbf{k}}$ accounts for independent band-to-band transitions. $\hat{v}$ is the bare electron–electron interaction, and $\hat{W}(\omega=0)$ is the statically screened interaction (see below). The BSE matrix becomes extremely large upon increasing the number of k-points and bands. The total size $N\times N$ can easily reach $N\approx 10^4 - 10^5$. Equation~\eqref{eq:exc} can be efficiently solved using different techniques discussed in the literature~\cite{Milev2026,Stohler2026}.

In the present work, the BSE analysis is performed on top of DFT+GW calculations for the determination of the quasi-particle spectrum. Ground-state (GS) calculations have been performed using the plane-wave Quantum Espresso density functional theory (DFT) code~\cite{Giannozzi2009}, within the Perdew-Burke-Ernzerhof (PBE) approximation for the exchange-correlation potential~\cite{Perdewetal1996}, and fully nonlocal SG15 ONCV two-projector norm-conserving Vanderbilt–Hamann pseudopotentials~\cite{Hamann2013}. Convergence tests, based on minimization of the GS total energy, yield converged results using a $12 \times 12 \times 1$ $k$-grid and an energy cutoff of 80\,Rydberg for the wave functions. Atomic positions have been relaxed to the PBE equilibrium lattice parameter of 5.901\,a.u..

For the optical spectrum, the BSE is solved on a $30\times 30 \times 1$ $k$-point grid, using 2 valence bands and 4 conduction states. For the XUV energy range, the BSE is solved on a $18\times 18 \times 1$ $k$-grid using the 8 semi-core states corresponding to Mo 3$p$ electrons, and a variable number of CBs up to 18 (see Sec.~\ref{Sec:NumRes}). From the solution of the eigenvalue problem, the polarizability $\tilde{\alpha}(\omega)$ [or dielectric function $\varepsilon(\omega)$] is reconstructed as a sum of excitonic poles weighted by excitonic dipoles.

To show the connection with the TD-HSEX scheme, we introduce the {\it ab initio} version of Eq.~\eqref{eq:hartree_exchange} for the Hartree potential and SEX self-energy in the KS basis set:
\begin{equation}
	\label{eq:hartree_sex_ai}
	\begin{aligned}
		v^{\mathrm{H}}_{nm\mathbf{k}}[\rho](t)  =\;&
		\sum_{n'm',\mathbf{k}'}
		v_{\substack{nm\mathbf{k}\\n'm'\mathbf{k}'}} \,
		\rho_{n'm',\mathbf{k}}(t) \\
		\Sigma^{\mathrm{SEX}}_{nm\mathbf{k}}[\rho](t) =\;&
		- \sum_{n'm',\mathbf{k}'}
		W_{\substack{nm\mathbf{k}\\n'm'\mathbf{k}}}(0)
		\, \rho_{n'm'\mathbf{k}'}(t)
	\end{aligned}
\end{equation}

The bare Coulomb interaction arises from the variation of the Hartree potential, ${\hat{v}=\partial \hat{v}^\mathrm{H}[\rho]/\partial \rho}$, whereas the screened interaction originates from the variation of the SEX self-energy, ${\hat{W}(0)=\partial\hat{\Sigma}^\mathrm{SEX}[\rho]/\partial\rho}$ (neglecting ${\partial\hat{W}(0)/\partial\rho}$). Within the two-particle framework, the Hartree kernel $\hat{v}$ describes an e–h exchange interaction, while the SEX kernel $\hat{W}(0)$ accounts for a direct e–h interaction responsible for binding.

The calculation of the screened interaction is one of the most demanding steps of the simulation. Compared to the case where a 2D RK potential is used, here the explicit calculation of screening fully accounts for the 2D nature of the material.

In \lumen the microscopic screening
\begin{equation}\label{eq:screening_g_space}
	\varepsilon^{-1}_{\mathbf{G,G}'}(\omega,\mathbf{q})=1+v_\mathbf{G}(\mathbf{q})\chi_{\mathbf{G,G}'}(\omega,\mathbf{q})
\end{equation}
is obtained through the solution of a Dyson equation for the density-density response function expressed in reciprocal space ($\mathbf{G}$-space) and at different transferred momenta $\mathbf{q}=\mathbf{k}-\mathbf{k}'$
\begin{multline}\label{eq:chi_g_space}
	\chi_{\mathbf{G,G}'}(\omega,\mathbf{q})=\chi^0_{\mathbf{G,G}'}(\omega,\mathbf{q})+ \\
	\chi^0_{\mathbf{G,G}''}(\omega,\mathbf{q})v_{\mathbf{G}''}(\mathbf{q})\chi_{\mathbf{G'',G'}}(\omega,\mathbf{q})\ .
\end{multline}
Here, the screening function $\varepsilon^{-1}_{\mathbf{G},\mathbf{G}'}(\omega,\mathbf{q})$ is computed within the plasmon-pole approximation, using a cutoff of 6\,Ha for the G expansion of the wave functions and 250 bands. The static part of $\hat{\varepsilon}^{-1}(\omega=0)$ is used to compute the effective potential $W$ in Eq.~\eqref{eq:exc}, via $\hat{W}(0)=\hat{\varepsilon}^{-1}(0)\hat{v}$.

To accelerate convergence, the approach is further refined using Coulomb-cutoff and random-integration techniques for both the bare and screened interaction~\cite{Guandalini2023}, which avoid spurious interactions between periodic replicas of the monolayer. As a by-product, by inverting Eq.~\eqref{eq:screening_g_space}, one can also obtain the macroscopic dielectric function or polarizability from the $\mathbf{G}=\mathbf{G}'=0$ components. The resulting optical properties differ from the BSE case in two main aspects: (i) only the exchange kernel $v$ enters the Dyson equation, and (ii) spectra can be efficiently computed including many conduction bands, since the matrix size $\chi_{\mathbf{G,G}'}(\mathbf{q},\omega)$ is independent of the number of bands. In the results section, we present absorption spectra including up to 250 bands in total (224 conduction and 36 valence bands), with a cutoff of 6\,Ha.

\subsection{On the role of screening}

By computing {\it ab initio} $\varepsilon^{-1}_{\mathbf{G,G}'}(\omega,\mathbf{q})$ via Eq.~\eqref{eq:screening_g_space}, screening is included only in the exchange term, while the Hartree contribution is evaluated using the bare interaction, as required by formal derivations. This is a key difference between an {\it ab initio} approach and the use of a 2D RK potential, which already embeds screening effects and material dimensionality. The use of a 2D RK potential has two consequences. On one hand (advantage), embedding the 2D geometry, $V^{\mathrm{RK}}(\mathbf{q})$ encodes the correct momentum dependence of the interaction in 2D materials, making convergence with respect to k-point sampling less demanding~\cite{Cudazzo2011}. On the other hand (drawback), it enforces the inclusion of screening effects in both Hartree and exchange terms. However, a screened Hartree term can be interpreted as an effective way to incorporate the role of high-energy CBs. Indeed, previous studies have shown that applying screening to the Hartree term in few-band calculations is equivalent to a down-folding approximation to calculations with many CBs and an unscreened Hartree term~\cite{Benedict2002,Deilmann2019}. This turns a main drawback of the RK potential into an advantage, which explains its widespread use.

In the {\it ab initio} framework, instead, both a large number of conduction bands and, especially in 2D materials, a very dense k-point sampling are required to converge optical properties~\cite{Huser2013}. The issue of k-point convergence has been addressed in several works~\cite{Huser2013,Qiu2016,Guandalini2023}. As discussed in the previous subsection, we adopt the solution proposed in Ref.~\cite{Guandalini2023}. The requirement of a large number of bands, however, remains, and convergence must be carefully checked.

In this work, we systematically investigate convergence with respect to the number of bands via {\it ab initio} calculations (\lumen code), and assess the validity of the down-folding approximation. In particular, we test brute-force convergence of the TD-Hartree case using the G-space implementation and the solution of Eq.~\eqref{eq:chi_g_space}. In parallel, we investigate the effect of varying the screening parameter in the RK potential (EDUS code). We find that both the number of bands and the screening parameter entering the Hartree term have a minor impact on the \mos absorption spectra in the IR/visible range [see Fig.~\ref{Fig:1}(c), dot-dashed curves]. In contrast, they significantly affect the spectral shape in the XUV regime, as shown in the following section.

\section{Numerical results and discussion}
\label{Sec:NumRes}
\begin{figure}[htbp]
	\begin{center}
		\begin{tabular}{c}
			\includegraphics[width=0.47\textwidth]{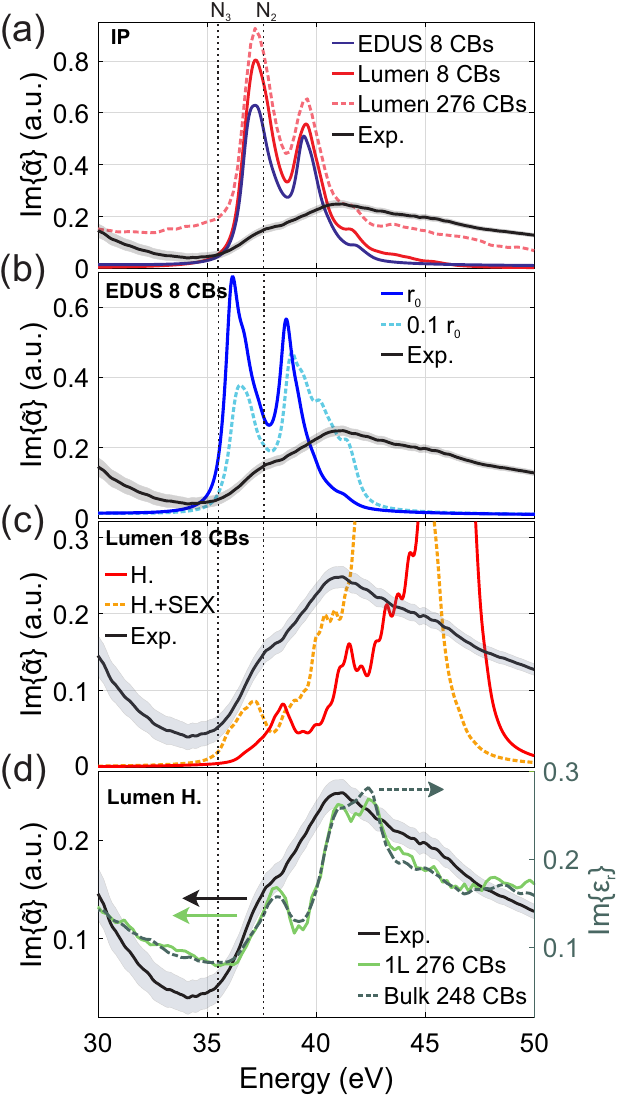}
		\end{tabular}
	\end{center}
	\caption[example]{\label{Fig:6} 
		Absorption spectra of 1L \mos at the Mo N$_{2,3}$ absorption edge. In all panels, the imaginary part of the complex polarizability, \ima, extracted from the experimental data is reported in solid black. The black shaded area corresponds to the estimated uncertainty (see Sec.~\ref{Sec:optconst}). \textbf{(a)} Comparison of IP spectra obtained with EDUS [dark blue, 8 conduction bands (CBs)] and \lumen (solid red, 8 CBs; dotted pink, 276 CBs - 300 in total). \textbf{(b)} \ima obtained with EDUS while considering 8 CBs, the SEX term, and employing an unscreened ($0.1r_0$, dotted cyan) or screened ($r_0$, solid blue) Hartree term. \textbf{(c)} \ima calculated with \lumen while considering 18 CBs and employing the Hartree term without (solid red) and with (dotted orange) the addition of the SEX term. \textbf{(d)} Absorption profile simulated with \lumen without the SEX term while considering 276 CBs for a 1L-\mos sample (solid light green) and 248 (300 in total) for a bulk crystal (dotted dark green). For the latter case, we plotted \ime to obtain a quantity independent of the crystal thickness. 
	}
\end{figure}

To assess the role of electronic correlations and explain the origin of the smooth resonance observed in the experimental data around the Mo N$_{2,3}$ edge [Fig.~\ref{Fig:4}(b)], we calculated the optical properties of 1L \mos with different levels of approximation and different numbers of CBs.

Three different levels of approximation are considered: (\textit{i}) Independent Particles (IP), where both the Hartree and exchange (SEX) terms are neglected; (\textit{ii}) Hartree, where the exchange term is neglected; and (\textit{iii}) the full HSEX solution. Here, through EDUS, we performed simulations with a different degree of screening of the Hartree term. 

The results are reported in Fig.~\ref{Fig:6}, which shows the imaginary part of the complex polarizability, \ima. Within the assumptions of this work, this quantity is directly proportional to the real part of the complex conductivity [see Eq.~\eqref{Eq:sig_alp}] and corresponds to the material absorption, providing a direct physical interpretation \cite{Tian2020}. In line with Fig.~\ref{Fig:5}, note that the experimental 1L-\mos \ima shown in all panels of Fig.~\ref{Fig:6} (black curves) corresponds to the 2L-\mos result rescaled by a factor of $\gamma = 0.5$ [see Eq.~\eqref{Eq:2L_1L}].

At first, we compare the simulation results obtained at the IP level in order to assess the role of electronic correlations. The results are shown in Fig.~\ref{Fig:6}(a), where the EDUS spectrum (dark blue, 8 CBs) has been rigidly red-shifted by 1.74\,eV to facilitate comparison with the \lumen output which includes QP corrections extracted from GW (solid red, 8 CBS; dotted pink 276 CBs). After this, theory and experiment show a reasonable energy alignment, and we do not introduce any additional energy shift. All IP approaches yield an absorption profile characterized by a pronounced peak at the Mo N$_{2,3}$ edge, which is absent in the experimental spectrum. This feature originates from direct transitions from core levels to the CBs. The missing background for the solid red and blue curves, i.e. vanishing \ima at low and high photon energies, reflects the restricted number of CBs, which truncates the available transition channels. Inserting all relevant transitions (dotted pink curve) the background is recovered. More importantly, the clear discrepancy between experiment and IP results in the Mo N$_{2,3}$ edge region highlights the role of electronic correlations, which are essential to accurately describe the optical response of TMDs.

To further investigate this aspect, we repeated the EDUS calculations including electron–electron correlations through both the Hartree and SEX terms. The results obtained using 6 core bands (corresponding to the Mo 4$p$ orbitals), 20 valence bands, and 8 CBs are reported in Fig.~\ref{Fig:6}(b), for both screened ($r_0 = 41.5$\,\AA, sold blue) and unscreened ($0.1r_0$, dotted cyan) Hartree term, using the same energy shift applied in Fig.~\ref{Fig:6}(a). Although both cases retain a peaked structure that does not match the experimental observation, likely due to the limited number of CBs, a particularly interesting effect emerges. While screening of the Hartree term is often introduced to mimic the contribution of electronic bands not explicitly included in the calculation \cite{Benedict2002,Deilmann2019}, here we find that better agreement with experiment is achieved in the absence of screening. In particular, the dotted cyan curve in Fig.~\ref{Fig:6}(b) yields a smoother and broader \ima, with a reduced blue shift, which more closely reproduces the experimental absorption profile. This result is unexpected, since the SEX term typically dominates excitations near the Fermi level (i.e., valence-to-CB transitions), whereas local-field (Hartree) effects are usually less important \cite{Wang2018}.

In Fig.~\ref{Fig:1}(c), we show three absorption profiles of 1L \mos in the IR/optical range, calculated with EDUS while progressively reducing the screening of the Hartree term (from $r_0$ to $0.01r_0$). In this low-energy regime, local-field effects are indeed modest, leading to an overall reduction of \ima with decreasing $r_0$, without significantly altering the spectral structure. Around the N$_{2,3}$ edge, however, a markedly different behavior emerges. When the Hartree term is screened, the calculated absorption profile becomes qualitatively similar to the IP result, although red-shifted by nearly 0.8\,eV [compare the dark-blue curve in Fig.~\ref{Fig:6}(a) with the blue curve in Fig.~\ref{Fig:6}(b)]. This observation highlights the important role of local-field effects in the XUV regime and indicates that their treatment requires particular care to avoid misinterpreting spectral features. Rather than promoting the formation of bound core excitons, our results suggest that correlations mediated by the Hartree term can substantially weaken, or even completely suppress, bound excitonic features associated with semi-core excitations.

To further assess the impact of the finite number of electronic bands and the relative role of Hartree and SEX, we perform many-body perturbation theory calculations with \lumen, since increasing the number of bands in real-time simulations such as EDUS is computationally prohibitive. Figure~\ref{Fig:6}(c) reports the absorption profiles of a \mos monolayer obtained with \lumen, including (H.+SEX, orange dotted curve) or excluding (H., solid red curve) the SEX term in addition to the Hartree contribution. In both cases, we considered the same core-hole decay time used in EDUS and included 18 CBs. As can be observed increasing the number of bands leads to a substantial renormalization of the absorption spectrum, with a redistribution of spectral weight toward higher photon energies that better resembles the experimental case compared to the IP prediction, nearby the N$_{2,3}$ edge. The inclusion of the SEX term (orange dotted curve) introduces a redshift of about 1.5\,eV, without qualitatively modifying this behavior. Notably, no sharp spectral features appear below the Mo N$_2$ and N$_3$ absorption edges.

This behavior, consistent with our experimental results, contrasts with W-based TMDs such as \ce{WS2} and \ce{WSe2}, where semi-core excitonic resonances produce sharp features below the absorption edges \cite{Chane2021,Oh2023}. The absence of such features in \mos supports the interpretation that the XUV optical response is dominated by collective transitions rather than bound semi-core excitons.

The redistribution of absorption intensity toward higher energies observed in the XUV regime is attributed to local-field (or depolarization) effects. These effects are known in molecules (0D systems \cite{Sottile2003,Onida2002}), in low-dimensional materials (1D and 2D) when probed along the non-periodic direction \cite{Bruno2005,Chang2004,Hong2016}, and in non-uniform 3D systems \cite{Sottile2003}. In \mos, the strong anisotropy between in-plane (ordinary axis) and out-of-plane (extraordinary axis) absorption \cite{Ermolaev2021} reflects the importance of local fields in the out-of-plane response. In the XUV regime, however, local-field effects also significantly influence in-plane absorption. This arises from the localized nature of semi-core hole excitations, which produce a non-uniform excitation density, combined with the relatively high polarizability of semi-core $p$ orbitals, in contrast to deeper core states and $d$ and $f$ semi-core edges (as observed in W-based compounds).

However, with 18 CBs an unphysical increasing of the signal above 42\,eV is observed. To accurately reproduce the experimentally observed spectral profile, it is then essential to include a substantially larger number of electronic bands. We therefore performed a second set of simulations with \lumen, incorporating all dipole-allowed transitions within the 30-50\,eV photon-energy window. In this case, the total number of bands increases to 300 (276 CBs). The results are shown in Fig.~\ref{Fig:6}(d). Remarkably, even without any energy shift or amplitude rescaling, the theoretical prediction (solid light-green curve) reproduces well the energy position, amplitude, and overall shape of the experimental edge (black curve with shaded area). A small blue shift is observed, likely due to the exclusion of the SEX term [compare with Fig.~\ref{Fig:6}(c)]. This level of agreement indicates that, in this energy range, the dominant contribution to the XUV response of \mos arises from single-particle transitions dressed by local-field effects, while excitonic correlations play only a minor role. Furthermore, the inclusion of a large number of CBs captures all main features of the spectrum between 30 and 50\,eV. There is a peak nearby the N$_3$ edge, corresponding to the shoulder seen in the experiment, and the spectral weight reaches a maximum around 41\,eV, as observed in the experimental data. Both features stem from collective excitations [they are not present at the IP level - pink dotted curve in Fig.~\ref{Fig:6}(a)] involving high energy conduction states which mix via local fields effects, similarly to what happens in plasmon resonances. Instead, including or excluding valence states in Figs.~\ref{Fig:6}(a)-(c) does not produce a visible effect.

Finally, we calculated the optical response of bulk \mos using the same theoretical framework, including 300 bands in total (which corresponds to 248 CBs), to assess the influence of dimensionality. The results are reported in Fig.~\ref{Fig:6}(d), right axis, where we show the imaginary part of the dielectric function, \ime (dotted dark-green curve), providing a quantity independent of thickness. Remarkably, almost no differences are observed. This is consistent with the predominantly atomic-like nature of the semi-core transitions governing the XUV response, and further supports the robustness of the interpretation across different dimensional regimes.

\section{Conclusions}

We reported multi-angle reflectivity measurements in the XUV spectral range for mono- and bilayer \mos samples on a \sn substrate, focusing on the energy region around the Mo N$_{2,3}$ and S L$_{2,3}$ edges. After characterizing the substrate optical response, we used a single-sheet 2D conductivity model to extract the complex optical properties of the \mos bilayer between 25 and 90\,eV. By introducing an effective thickness equal to the interlayer spacing ($d = 0.650$\,nm), we derived an effective complex refractive index $\tilde{n}$, which was then used within a 3D slab model to reproduce the measured reflectivity, demonstrating consistency between the two descriptions.

Extending the analysis to the monolayer limit, we find that the experimental reflectivity is equivalently reproduced either by a 3D model using the same $\tilde{n}$ with halved thickness, or by a 2D model with the conductivity reduced by a factor of two. Within experimental uncertainty, this indicates a linear scaling of the optical response with thickness, supporting the applicability of both representations in this energy range.

The extracted absorption-related quantities, namely the real part of the 2D conductivity and the imaginary part of the refractive index or polarizability, display a broad resonance at the Mo N$_{2,3}$ edge without signatures of sharp excitonic features. This behavior is consistent with previous reports on bulk Mo-based TMDs and contrasts with observations in W-based compounds. These results indicate that excitonic effects play a reduced role in shaping the XUV optical response of atomically thin \mos, despite their dominant role in the visible and near-IR regimes.

To rationalize these findings, we performed first-principles simulations of the monolayer absorption spectrum using both EDUS and \lumen. The comparison across different levels of approximation and increasing numbers of electronic sub-bands shows that local-field effects associated with the Hartree term dominate the response in this energy range. In contrast, screened-exchange (SEX) contributions, while essential for exciton formation at lower energies, primarily induce rigid energy shifts and do not qualitatively modify the spectral profile. We further find that, unlike in the visible regime, reducing the Hartree screening to compensate for a limited number of sub-bands leads to moderate smoothing and broadening, but remains insufficient to reproduce the experimental spectrum. Quantitative agreement requires the inclusion of a substantially larger number of conduction bands.

Overall, our results demonstrate that, even in the monolayer limit where reduced dielectric screening typically enhances excitonic effects, the XUV response of \mos is governed predominantly by local-field effects associated with the Hartree interaction. This suggests that the suppression of excitonic signatures at high photon energies is not solely driven by dimensionality and may persist in the bulk. Consistently, \lumen calculations for bulk \mos, including a large manifold of electronic states, yield a smooth absorption profile in agreement with both experiment and existing literature.

Finally, these findings highlight that the role of reduced dimensionality in core-level excitations differs fundamentally from that governing band-edge excitons. As a result, concepts developed for the visible and near-IR response of TMDs cannot be directly transferred to the XUV regime. This has implications for the interpretation of attosecond spectroscopies and for the design of field-driven experiments, where the choice of transition metal and the nature of the probed electronic states must be carefully considered.

\begin{acknowledgments}
	M.L. and N.D.P. wish to thank Prof. Andrea Marini for fruitful discussion. The experiments were carried out at the BEAR beamline at Elettra (proposal 20240231). This project has received funding from the European Research Council (ERC) under the European Union’s Horizon 2020 research and innovation programme (grant agreement No.~848411 title AuDACE), from MUR FARE (Grant No.~R209LXZRSL, title PHorTUNA) and the MUR PRIN 2022 – Scorrimento (Grant ID~2022PX279E, title exATTO). D.S. aknoledges funding from the European Union’s Horizon Europe research and innovation programme under the Marie Sklodowska-Curie grant agreement 101118915 (project TIMES) and the MaX "MAterials design at the eXascale” co-funded by the European High Performance Computing joint Undertaking (JU) and participating countries (Grant Agreement No. 101093374).
	A.P. acknowledges the Spanish Ministry of Science, Innovation and Universities \& the State Research Agency through grants refs. PID2024-157663NB-I00 (MCIU/AEI/FEDER, UE), and the ``Severo Ochoa" Programme for Units of Excellence in R\&D (CEX2024-001445-S), and computer resources and assistance provided by the Centro de Computaci\'on Cient\'ifica (CCC-UAM) and the Red Espa\~nola de Supercomputaci\'on (RES) under projects refs. FI-2025-2-0036, FI-2024-3-0011, and FI-2024-2-0034. M.M. thanks the Ministry of Science and Higher Education of the Russian Federation for supporting theoretical calculations through funding the Institute of Metal Physics, as well as the Uran supercomputer at the IMM UB RAS for providing computational resources. Authors also acknowledge the CINECA award under the ISCRA initiative, for the availability of high performance computing resources and support.
\end{acknowledgments}

\bibliography{references}

\end{document}